\documentclass[pss]{wiley2sp} 
\usepackage{amsmath}
\usepackage{verbatim}

\begin{document}

\title{Quantum entanglement of space domains occupied by interacting electrons}

\titlerunning{Quantum entanglement of space domains}

\author{%
  A. Ram\v{s}ak\textsuperscript{\textsf{\bfseries 1,2\Ast}},
  J. Mravlje\textsuperscript{\textsf{\bfseries 1}},
  T. Rejec\textsuperscript{\textsf{\bfseries 1,2}}
}
\authorrunning{A. Ram\v{s}ak et al.}

\mail{e-mail
  \textsf{anton.ramsak@fmf.uni-lj.si}}

\institute{%
   \textsuperscript{1}\,Faculty of Mathematics and Physics, University
  of Ljubljana, Jadranska 19, 1000 Ljubljana, Slovenija\\
 \textsuperscript{2}\,J. Stefan Institute, Jamova 39, 1000
  Ljubljana, Slovenija\\
}

\received{XXXX, revised XXXX, accepted XXXX} 
\published{XXXX} 



\abstract{%
%
%
%
  Spin-entanglement of two electrons occupying two spatial regions --
  domains -- is expressed in a compact form in terms of spin-spin
  correlation functions. The power of the formalism is demonstrated on
  several examples ranging from generation of entanglement by
  scattering of two electrons to the entanglement of a pair of qubits
  represented by a double quantum dot coupled to leads. In the latter
  case the collapse of entanglement due to the Kondo effect is
  analyzed.  }{%
}

%
%

\maketitle   

\section{Introduction}\label{sec1}

Based on the peculiar behaviour of entangled quantum states, in 1935
Einstein, Podolsky and Rosen argued that quantum mechanical
description of reality is not complete \cite{einstein35}.  Today
this article is  Einstein's most cited publication, but quantum
entanglement is not considered a paradox. In fact, the ability to
establish entanglement between quantum particles in a controlled
manner is a crucial ingredient of any quantum information processing
system \cite{nielsen01}. Also, the study of entanglement provides
insight into the nature of many-body states in the vicinity of
crossovers between various regimes or points of quantum phase
transition \cite{osterloh02}.

In realistic hardware designed for quantum information processing,
several criteria for qubits (DiVincenzo's checklist) must be fulfilled
\cite{criteria}: the existence of multiple identifiable qubits, the
ability to initialize and manipulate qubits, small decoherence, and
the ability to measure qubits, {\it i.e.}, to determine the outcome of
computation. Quite generally, the parts of an interacting system are
to some extent entangled.  However, fully entangled qubit pairs are
required for such applications \cite{nielsen01}.  This leads to the
question of how to quantify the entanglement.

The entanglement of binary quantum objects (qubits) can be quantified
by the entanglement of formation, a notion which for
pure states reduces to the von Neumann entropy
\cite{bennett96,hill97,vedral97}. This measure is convenient since it
is proportional to the number of fully entangled pairs needed to
produce a given entangled state. Also, it is monotonically related to
the concurrence, which can be evaluated by the Wootters formula
\cite{wootters98} as a simple function of the density matrix.

A possible realization of a qubit pair can be achieved by confining
electrons to two spatial regions. In this case one has the freedom to
choose between spin or charge of the electron to represent a
qubit. The entanglement measures must account for both
possibilities. Additional complications arise due to the
indistinguishability of the particles and possible states of multiple
occupancy \cite{schliemann01,ghirardi04,zanardi02,vedral03}.

On the experimental side, it seems that among several proposals for
quantum information processing systems, the criteria for scalable
qubits can be met in solid state structures consisting of coupled
quantum dots \cite{divincenzo05,coish06}. The ability to precisely
control the number of electrons \cite{elzerman03} and the evidence
for spin entangled states have been reported in GaAs based
heterostructures \cite{chen04,hatano05}. It has also been
demonstrated that in double quantum dot systems coherent qubit
manipulation and projective readout are possible \cite{petta05}.

\begin{figure}
\begin{center}\includegraphics[%
  width=55mm,
  keepaspectratio]{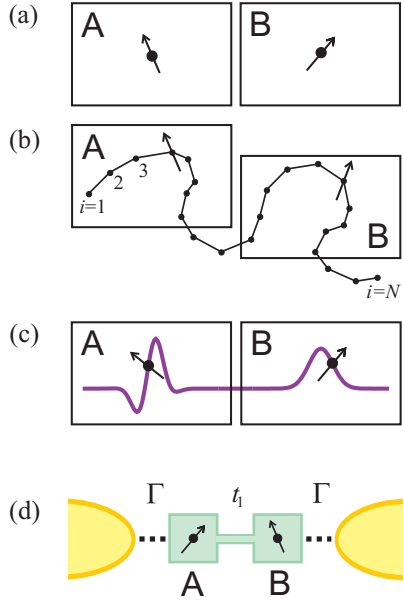}
 
\end{center}
\caption{\label{fig1}(Color online) (a) Two electrons (qubits) localized in separate quantum dots A
and B. (b) Two delocalized electrons in separate measurement domains A
and B; the probability of finding one electron in a domain is equal for A
and B: $n_{A}=n_{B}=1$. (c) Scattering of two electrons.
(d) Double quantum dot system coupled to the leads: the chemical potential is set such that each of the dots is occupied by one interacting electron  on average.}
\end{figure}

Here we explicitely evaluate the concurrence in terms of spin-spin
correlation functions between the two subsystems. In this manner we
easily account for entanglement between subsystems containing
delocalized electrons. We demonstrate the convenience of our approach
by evaluating the time-dependence of concurrence in a toy example of
two electrons in a three site chain and in a more realistic entangler
based on scattering of electrons in a quantum wire.

On the other hand, entangled qubits can also be extracted from
interacting many-body systems. Our formulas are readily generalized
also for this case and even for finite temperature. As an example, we
discuss the case of two quantum dots coupled serially to metallic
electrodes. In this case, entanglement of a pair of electrons that
are confined in a double quantum dot may collapse at low temperatures
due to the Kondo effect even for a very weak coupling to the leads.

The paper is organized as follows.  Sec. \ref{sec2} introduces the
entanglement measure for two delocalized electrons, which can be
simplified for some special cases elaborated in Sec.~\ref{sec3}. In
Sec.~\ref{sec4} we illustrate the convenience of our approach with
numerical examples mentioned above. Conclusions are given in
Sec.~\ref{sec5}.

\section{Entanglement of two delocalized electrons}\label{sec2}

A pure state of two spins represented by electrons, each occupying a
separate quantum dot $A$ and $B$, can be described in the basis of
single spin-$\frac{1}{2}$ states $s=\uparrow$ or $\downarrow$ as
$|\Psi_{AB}\rangle=\sum_{ss'}\alpha_{ss'}|s\rangle_{\!
A}|s'\rangle_{\! B}$. Such a state, Fig.~\ref{fig1}(a), is
factorizable $|\Psi_{AB}\rangle=|\phi_{A}\rangle |\phi_{B}\rangle$ if
and only if
$\alpha_{\uparrow\!\downarrow}\alpha_{\downarrow\!\uparrow}=\alpha_{\uparrow\!\uparrow}\alpha_{\downarrow\!\downarrow}$
as easily checked by explicit construction. For general coefficients
$\alpha_{ss'}$ the state is not factorizable and its entanglement can
be quantified by concurrence $C$ as introduced by Hill and Wootters
\cite{hill97},

\begin{equation}
C =  2|\alpha_{\uparrow\!\downarrow}\alpha_{\downarrow\!\uparrow}-\alpha_{\uparrow\!\uparrow}\alpha_{\downarrow\!\downarrow}|.\label{eq:wooters}
\end{equation}

Consider now a more general problem of two electrons in a state where
the system cannot directly be reduced to an equivalent system with
(pseudo) spin degrees of freedom only. Let us take for example two
electrons on a lattice \cite{ramsak06} described by a state
\begin{equation}
|\Psi\rangle=\sum_{(i,j) \; ss'}\psi_{ij}^{ss'}c_{i s}^{\dagger}c_{j s'}^{\dagger}|0\rangle ,\label{eq:psi}\end{equation}
where $c_{is}^{\dagger}$ creates an electron with spin $s$ on site $i$
on the lattice with the total number of sites $N$. Sites of the
lattice are assumed to be ordered as for example in Fig.~\ref{fig1}(b)
and $(i,j)$ corresponds to the summation over all sites $i=1, ... N$
and $i \le j$ \cite{notedifference}.

Such states may arise when two initially unentangled electrons in wave
packets approach each other, then interact and finally become again
well separated in distinct regions A and B,
Fig.~\ref{fig1}(c), where they could be extracted for further
purposes. Alternatively, such states can be realized in various
correlated electron systems and, in order to study them theoretically,
elaborate many-body techniques are sometimes needed. Moreover, usually
not the state itself but only the correlation functions are available.

Therefore it is advantageous to express the entanglement in terms of
spin-spin correlation functions. The spin operators for a single electron
occupying domain A (or B) are expressed as the sum of operators for
sites $i$ within the given domain, 
\begin{equation}
S_{A(B)}^{\lambda}=\frac{1}{2}\sum_{i\in A(B)}\sum_{ss'}c_{is}^{\dagger}\sigma_{ss'}^{\lambda}c_{is'},
\end{equation}
where $\sigma^{\lambda}_{ss'}$ are the Pauli matrices.

The concurrence is given by the eigenvalues of the non-Hermitian
matrix $\rho\tilde{\rho}$, where $\rho$ and $\tilde{\rho}$ are the
reduced density matrix and its time reverse, respectively
\cite{wootters98}. For axially symmetric problems, where both
$\langle\Psi|S_{A(B)}^{x}|\Psi\rangle=0$ and
$\langle\Psi|S_{A(B)}^{y}|\Psi\rangle=0$ as well as
$\langle\Psi|S_{A(B)}^{z}S_{B(A)}^{x}|\Psi\rangle=0$ and
$\langle\Psi|S_{A(B)}^{z}S_{B(A)}^{y}|\Psi\rangle=0$, it can be
expressed in a compact form \cite{ramsak06}
\begin{eqnarray}
C & = & \textrm{max}(0,C_{\uparrow\!\downarrow},C_{\parallel}),\label{eq:cmax}\\
C_{\uparrow\!\downarrow} & = & 2|\langle S_{A}^{+}S_{B}^{-}\rangle|-2\sqrt{\langle P_{A}^{\uparrow}P_{B}^{\uparrow}\rangle\langle P_{A}^{\downarrow}P_{B}^{\downarrow}\rangle},\nonumber \\
C_{\parallel} & = & 2|\langle S_{A}^{+}S_{B}^{+}\rangle|-2\langle P_{A}^{\uparrow}P_{B}^{\downarrow}\rangle,\nonumber 
\end{eqnarray}
where $S_{A(B)}^{+}=(S_{A(B)}^{-})^{\dagger}=\sum_{i\in A(B)}c_{i\uparrow}^{\dagger}c_{i\downarrow}$
are spin raising operators for domains A (or B) and 
\begin{equation}
P_{A(B)}^{s}=\sum_{i\in A(B)}n_{is}(1-n_{i,-s})
\end{equation}
are spin-$s$ projectors operating
in domains A (or B) with $n_{is}=c_{is}^{\dagger}c_{is}$. Fermionic expectation
values required in Eq.~(\ref{eq:cmax}) are then given as
\begin{eqnarray}
\langle S_{A}^{+}S_{B}^{-}\rangle & = & \sum_{[ij]}\psi_{ij}^{\uparrow\!\downarrow*}\psi_{ij}^{\downarrow\!\uparrow},\nonumber \\
\langle S_{A}^{+}S_{B}^{+}\rangle & = & \sum_{[ij]}\psi_{ij}^{\uparrow\!\uparrow*}\psi_{ij}^{\downarrow\!\downarrow},\label{eq:sasb}\\
\langle P_{A}^{s}P_{B}^{s'}\rangle & = & \sum_{[ij]}|\psi_{ij}^{ss'}|^{2},\nonumber \end{eqnarray}
where $[ij]$ in Eqs.~(\ref{eq:sasb}) corresponds to the summation over
all pairs $i,j$ such that $i\in A$ and $j\in B$. Concurrence formula
Eq.~(\ref{eq:cmax}) is valid as long as double occupancy of sites is
negligible, $\langle n_{i s} n_{i s'} \rangle \to 0$.  It is assumed
that the wave function is normalized,
\begin{equation}
\langle\Psi |\Psi\rangle= \sum_{(i,j),\; ss' }|\psi_{ij}^{ss'}|^{2}=1.
\end{equation}
We stress that the correlation functions in Eq.~(\ref{eq:cmax}) can be
evaluated for pure or mixed states, the latter arising, for instance,
due to the finite temperature or by tracing out the environment
degrees of freedom, {\it e.g.}, leads in Fig.~1(d).
\section{Special cases}\label{sec3}

In states with the SU(2) symmetry $\langle
S_{A}^{x}S_{B}^{x}\rangle=\langle S_{A}^{y}S_{B}^{y}\rangle=\langle
S_{A}^{z}S_{B}^{z}\rangle$ and the concurrence formula
Eq.~(\ref{eq:cmax}) simplifies further to a function depending on only
one spin invariant $\langle\textrm{\bf S}_{\mathrm{A}}\cdot\textrm{\bf
S}_{\mathrm{B}}\rangle$,
\begin{equation}
C_{\mathrm{A}  \mathrm{B}  } = \textrm{max}\left(0,-2\langle\textrm{\bf
S}_{\mathrm{A}}\cdot\textrm{\bf
S}_{\mathrm{B}}\rangle-\frac{1}{2}\right).
\label{su2}
\end{equation}
The concurrence is expected to be significant whenever enhanced
spin-spin correlations indicate A-B singlet formation.

If $|\Psi\rangle$ is an eigenstate of the total spin projection
$S_{\textrm{tot}}^{z}$ the concurrence is given solely with the overlap between $|\Psi\rangle$
and the $AB$-spin-flipped state $|\tilde{\Psi}\rangle=|S_{A}^{+}S_{B}^{-}\Psi\rangle$. If
$S_{\textrm{tot}}^{z}=\pm1$ the concurrence is zero, while for $S_{\textrm{tot}}^{z}=0$
\begin{equation}
C =  C_{\uparrow\!\downarrow}=2|\sum_{[ij]}\psi_{ij}^{\uparrow\!\downarrow*}\psi_{ij}^{\downarrow\!\uparrow}|,\label{eq:cpsi}\end{equation}
which is a generalization of the concurrence formula Eq.~(\ref{eq:wooters}) to $N$ sites.

The concurrence formulas, Eq.~(\ref{eq:cmax}), remain essentially the
same if the state $|\Psi\rangle$ corresponds to the system in
continuum space, $i\rightarrow{\textbf{r}}=(x,y,z)$, the only change
being integrations over the corresponding measurement domains,
\begin{equation}
C=|\int_{A}\!\int_{B}\!(\varphi_0\!-\varphi_1)^{*}(\varphi_0\!+\varphi_1)\textrm{d}^{3}{\textbf{r}}_{1}\textrm{d}^{3}{\textbf{r}}_{2}|,\label{eq:xyz}\end{equation}
where $\varphi_S\equiv \varphi_S(\textbf{r}_{1},{\textbf{r}_2)=\langle{\textbf{r}}_{1},{\textbf{r}}_{2};S}|\Psi\rangle$ are singlet and triplet amplitudes for $S=0$ and $S=1$, respectively.

Another interesting special case is the wave function $|\Psi\rangle$ which is a linear combination of  entangled Bell $AB$-pairs,
\begin{equation}
|\Psi\rangle=\sum_{\beta=1}^{4}b_{\beta}\sum_{[ij]}\psi_{ij}|ij,\beta\rangle,\,\,\,\,\,\sum_{[ij]}|\psi_{ij}|^{2}=1,\label{eq:c1}
\end{equation}
where for each  pair of sites $(i,j)$ one can introduce the Bell
basis $|ij,\beta\rangle$ \cite{bennett96},
\begin{eqnarray}
|ij,1\rangle&=&\frac{1}{\sqrt{2}}(c_{i\uparrow}^{\dagger}c_{j\uparrow}^{\dagger}+c_{i\downarrow}^{\dagger}c_{j\downarrow}^{\dagger})|0\rangle,\\ \nonumber
|ij,2\rangle&=&\frac{\imath}{\sqrt{2}}(c_{i\uparrow}^{\dagger}c_{j\uparrow}^{\dagger}-c_{i\downarrow}^{\dagger}c_{j\downarrow}^{\dagger})|0\rangle,\\ \nonumber
|ij,3\rangle&=&\frac{\imath}{\sqrt{2}}(c_{i\uparrow}^{\dagger}c_{j\downarrow}^{\dagger}+c_{i\downarrow}^{\dagger}c_{j\uparrow}^{\dagger})|0\rangle,\\ \nonumber
|ij,4\rangle&=&\frac{1}{\sqrt{2}}(c_{i\uparrow}^{\dagger}c_{j\downarrow}^{\dagger}-c_{i\downarrow}^{\dagger}c_{j\uparrow}^{\dagger})|0\rangle.\\
\end{eqnarray}
In this case, the concurrence is given with a simple expression $C=|\sum_{\beta}b_{\beta}^{2}|$.

\section{Numerical examples}\label{sec4}

Here we use concurrence formulas in practice. We evaluate the
concurrence for a few examples of interacting electrons on a lattice
described by the following generic hamiltonian
\begin{equation}
H=-\sum_{ijs}(t_{ij}c_{is}^{\dagger}c_{js}+h.c.)+\sum_{ijss'}U_{ij}n_{is}n_{js'}.\label{masterh}
\end{equation}
For simplicity we take the electron-electron interaction constant up to some distance, {\it i.e.},
$U_{ij}=\frac{1}{2}U\sum_{m=0}^{M}\delta_{|i-j|,m}$.
\begin{figure}
\begin{center}\includegraphics[%
  width=65mm,
  keepaspectratio]{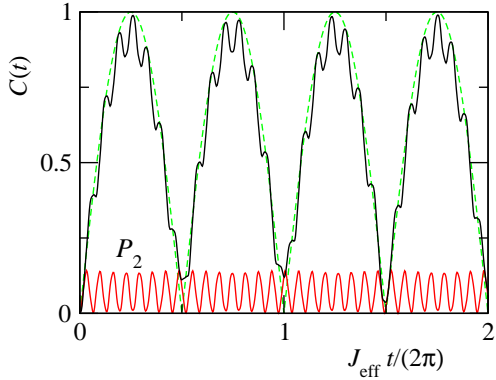}
\end{center}
\caption{\label{fig2}(Color online) Full line represents concurrence $C(t)$ for a three sites system with $t_{12}/t_{23}=1/10$, $U/t_{23}=5$ and $J_\textrm{eff}/t_{23}=0.023$ (singlet-triplet energy difference). Dashed line corresponds to the strong coupling limit results $C_\textrm{eff}(t)$ with the same $J_\textrm{eff}$. Domains are doubly occupied with the probability $P_2$.}
\end{figure}

\subsection{Two qubits on three sites} 
First we consider two electrons on three sites, where site $i=1$
corresponds to the measuring domain A and sites $i=2,3$ to the
domain B. We take the Hubbard model, $M=0$, and two non-zero
hopping matrix elements, $t_{12}$ and $t_{23}$. In the limit
$t_{23}>t_{12}$ a bonding orbital is formed between the sites 2 and 3
and in the ground state of the system there is a single electron in
each of the domains. For large $U$, the ground state is a spin singlet
formed between site $i=1$ and the bonding orbital, with the excitation
energy $J$ to the triplet state.

If in each of
the domains there is precisely one electron and the state is an
eigenstate of total spin projection, $S_{\textrm{tot}}^{z}=0$,
Eq.~(\ref{eq:cmax}) simplifies to
\begin{equation}
C =2|\psi_{12}^{\uparrow\!\downarrow*}\psi_{12}^{\downarrow\!\uparrow}+\psi_{13}^{\uparrow\!\downarrow*}\psi_{13}^{\downarrow\!\uparrow}|.\label{eq:cpsi3}
\end{equation}

Let us put the electrons to the system in an initially separable state
consisting of a spin up electron in A and the other electron with
spin down in the bonding orbital of B,
$|\Psi(0)\rangle=c^\dagger_{1\uparrow}{1 \over \sqrt{2}}(
c^\dagger_{2\downarrow}+c^\dagger_{3\downarrow})|0\rangle$. Because
the inital state is composed of different energy eigen-states, the
Rabi oscillations occur. In the strong coupling limit, $U,t_{23} \gg
t_{12}$ the system is described by the Heisenberg model with
antiferromagnetic coupling $J_\textrm{eff}\sim8t_{12}^2/(U - 2
t_{23})$ between the site 1 and the bonding orbital. In this limit the
Rabi oscillations occur due to the singlet-triplet splitting and the
time evolution of concurrence is given by $C_\textrm{eff}=|\sin
J_\textrm{eff} t|$.

For generic values of parameters additional states, for example the
states when the site 1 is doubly occupied, become relevant. We show
the concurrence for such a case in Fig.~\ref{fig2} and compare it to
the simplified expression given above. The simple behaviour is partly
reproduced but additional oscilations arise on other time scales.  We
plot also the probability that both electrons simultaneously occupy
one of the domains, $P_2=1- \langle
n_{1\uparrow}+n_{1\downarrow}\rangle+ 2 \langle
n_{1\uparrow}n_{1\downarrow}\rangle$. The oscillations of this
quantity occur on the time scale given by the characteristic time of
tunneling events, $t_{12}^{-1}$.

\subsection{Two flying qubits} 

\begin{figure}
\begin{center}\includegraphics[%
  width=65mm,
  keepaspectratio]{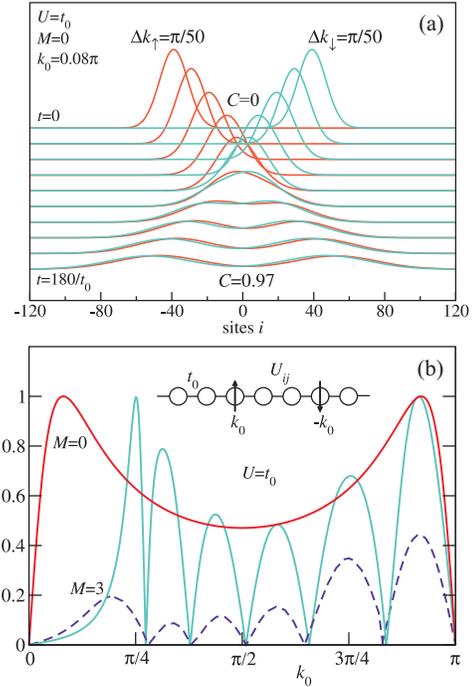}\end{center}

\caption{\label{fig3}(Color online) (a) Time evolution of spin $\uparrow$
and $\downarrow$ electron density during the scattering for the Hubbard model $(M=0)$
with $U=t_{0}$ and $\Delta k=\pi/50$ and $k_0$ near the concurrence maximum. 
(b) $C$ as a function of $k_0 $: (i) the Hubbard model
$(M=0)$ with $U=t_{0}$ and $\Delta k=0$; (ii) $M=3$: with $\Delta k=0$
(full line), and (iii) $\Delta k=\pi/10$ (dashed). }
\end{figure}

Next we consider two flying qubits, {\it i.e.}, two electrons on an
infinite one-dimensional lattice with the hamiltonian
Eq.~(\ref{masterh}) and $t_{ij}=t_0$ for $j=i+1$.  To be specific, let
one electron with spin $\uparrow$ be confined initially to region $A$
($i\sim-L$) and the other electron with opposite spin to region $B$
($i\sim L)$, Fig.~\ref{fig3}. The simplest initial state consists of
two wave packets with vanishing momentum uncertainties $\Delta
k\rightarrow0$, with momenta $k>0$ and $q<0$ for the left and the
right wave packet, respectively. After the collision, the electrons
move apart with a probability amplitude ${\cal M}_{| |}(k,q)$ for
non-spin-flip scattering and a spin-flip amplitude ${\cal
M}_{\uparrow\!\downarrow}(k,q)$. Concurrence after the collision is
then readily expressed from Eq.~(\ref{eq:cpsi}) as
\begin{equation}
C=2|{\cal M}_{| |}(k,q){\cal M}_{\uparrow\!\downarrow}(k,q)|.
\end{equation}
Note that $C=1$ when non-spin-flip and spin-flip amplitudes coincide
in accord with recent analysis of flying and static qubits
entanglement \cite{jrr05,gunlycke05,giavaras06,wires}.

More general initial wave packets with finite $\Delta k$ are defined
with appropriate momentum amplitudes $\phi_{\uparrow}(k)$ and
$\phi_{\downarrow}(q)$ for spin $\uparrow$ and $\downarrow$,
respectively. The concurrence as follows from Eq.~(\ref{eq:cpsi})
\begin{equation}
C=2|\int\!\!\!\int {\cal M}_{| |}^*(k,q){\cal M}_{\uparrow\!\downarrow}(k,q)
|\phi_{\uparrow}(k)\phi_{\downarrow}(q)|^{2}\textrm{d}k\textrm{d}q|\label{eq:ckq}\end{equation}
consists of a coherent superposition of scattering amplitudes
\cite{ramsak06}.

\begin{figure}
\begin{center}\includegraphics[%
  width=65mm,
  keepaspectratio]{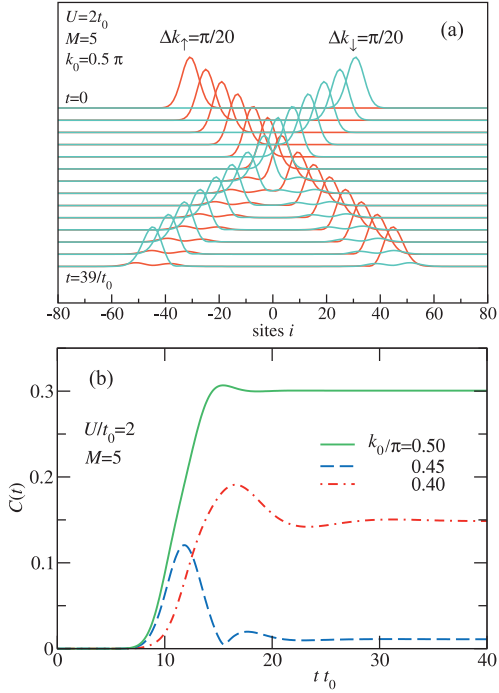}\end{center}

\caption{\label{fig4}(Color online) (a) Time evolution of spin
$\uparrow$ and $\downarrow$ electron density for $U=2t_{0}$, $M=5$
with $k_{0}=0.5\pi$, $\Delta k_{\uparrow}=\Delta
k_{\downarrow}=\pi/20$.  (b) $C(t)$ for Gaussian packets with various
$k_{0}$, $M=5$, $U=2t_{0}$ and $\Delta k=\pi/20$.  At $t=0$ the
separation between the packets is $2L=10/\Delta k$.}
\end{figure}

\begin{figure}
\begin{center}\includegraphics[%
  width=65mm,
  keepaspectratio]{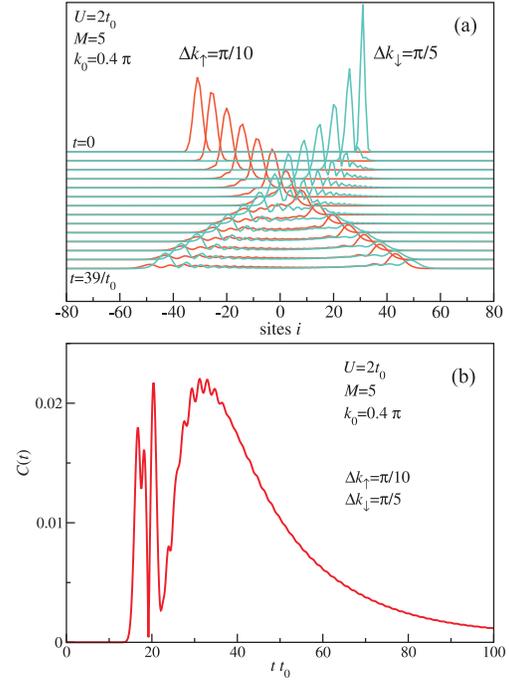}\end{center}

\caption{\label{fig5}(Color online) As in Fig.~\ref{fig4}, with parameters 
$U=2t_{0}$, $M=5$ but different
$k_{0}=0.4\pi$, $\Delta k_{\uparrow}=\pi/10$ and $\Delta k_{\downarrow}=\pi/5$.}
\end{figure}

The simplest example is the Hubbard model where the scattering
amplitudes can be obtained analytically for the case of one-dimension,
${\cal M}_{| |}(k,q)=1+{\cal M}_{\uparrow\!\downarrow}(k,q)=(\sin
k-\sin q)/[\sin k-\sin q+\imath U/(2t_{0})]$ \cite{lieb}.  In
Fig.~\ref{fig3}(a) the time evolution of spin $\uparrow$ and
$\downarrow$ electron densities is presented. In Fig.~\ref{fig3}(b)
the corresponding concurrence for wave packets with a well defined
momentum $k_{0}$ for $U=t_{0}$ is shown, together with a longer range
interaction case, $M=3$, for a sharp momentum (full line) and for a
Gaussian initial amplitude $\phi_{\downarrow}(q)=\phi_{\uparrow}(-k)$
with $\Delta k=\pi/10$ (dashed line). An interesting observation here
is a substantial reduction of the concurrence due to coherent
averaging in Eq.~(\ref{eq:ckq}). Additionally, electrons will be
completely entangled at some kinetic energy comparable with the
repulsion, $U\sim2t_{0}(1-\cos k_{0})$, where non-spin-flip and
spin-flip amplitudes coincide. In Fig.~\ref{fig4}(a) and
Fig.~\ref{fig5}(a) some representative additional examples of time
evolution of interacting wave packets are shown.

The concurrence formula Eq.~(\ref{eq:cpsi}) is derived for electronic
states when double occupancy is negligible, which in our case is
strictly fulfilled only asymptotically when the electrons are far
apart. However, Eq.~(\ref{eq:cpsi}) can be evaluated at any time $t$
and the resulting $C(t)$ can serve as a measure of entanglement during
the transition from initial to final state.  In Fig.~\ref{fig4}(b) and
Fig.~\ref{fig5}(b) $C(t)$ corresponding to parameters of
Fig.~\ref{fig4}(a) and Fig.~\ref{fig5}(a) is shown. Concurrence
oscillations can be interpreted as a response to the finite time
duration of electron-electron interaction -- exchange -- where the
model can be approximately mapped onto an effective Heisenberg model
as in the case of three sites presented above.

\subsection{Qubit pairs in coupled quantum dots}

One of the simplest realizations of a solid state qubit is a serially
coupled double quantum (DQD). We model such a DQD using the
two-impurity Anderson Hamiltonian
\begin{eqnarray}
H&=&
\sum_{i=\mathrm{A},\mathrm{B}}(\epsilon n_{i}+Un_{i\uparrow}n_{i\downarrow})
+ V n_{\mathrm{A}}n_{\mathrm{B}}\\ \nonumber
& &-t_1\sum_{s}(c_{\mathrm{A}s}^{\dagger}c_{\mathrm{B}s} + h.c.),
\label{2and}
\end{eqnarray}
where $c^\dagger_{is}$ creates an electron with spin $s$ in the dot
$i=\mathrm{A}$ or $i=\mathrm{B}$ and $n_{is}=c^\dagger_{is}c_{is}$ is
the number operator. The on-site energies $\epsilon$ and the Hubbard
repulsion $U$ are taken equal for both dots. The dots are coupled to
the left and right noninteracting tight-binding leads with the
chemical potential set to the middle of the band of width $4t_0$. Each
of the dots is coupled to the adjacent lead by hopping $t_2$ and the
corresponding hybridization width is
$\Gamma=(t_2)^2/t_0$. Schematically this setup is presented in
Fig.~\ref{fig1}(d). The dots are additionally coupled capacitively by
a inter-dot repulsion term $V n_{\mathrm{A}}n_{\mathrm{B}}$.

\begin{figure}

\center{\resizebox{0.85\columnwidth}{!}{%
\includegraphics{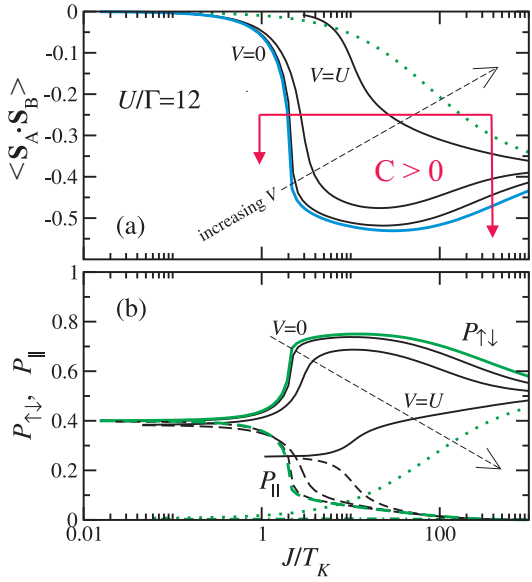} }}
\caption{(a) Spin-spin correlation for $V/U=0,1/3,2/3,1$ (full lines)
  and $V/U=5/4$ (dotted) for $U/\Gamma=12, \Gamma/t_0=0.1$. (b)
  Probabilities for parallel (lower curves - dashed) and anti-parallel
  (upper curves - full) spins of electrons in the DQD for $V/U$ ratios
  as in (a).  Note that the probability for parallel spins for
  $V/U=5/4$ is almost zero (dashed-dotted), while
  $P_{\uparrow\downarrow}<1/2$ for $J/T_K<1000$ (dotted); the
  probabilities do not sum to 1. The deficiency (which goes to zero as
  $U\to \infty$) is due to the states with double occupancy on at
  least one of the dots. }
\label{fig6}       
\end{figure}

When a DQD is attached to leads the low temperature physics is to a
large extent the same as that of the two-impurity Kondo problem
\cite{jones89}. There two impurities form either two Kondo singlets
with delocalized electrons or bind into a local spin-singlet state
which is virtually decoupled from delocalized electrons. The crossover
between the regimes is determined by the relative values of the
exchange magnetic energy $J$ and twice the Kondo condensation energy,
of order the Kondo temperature given by the Haldane formula,
$T_K=\sqrt{U\Gamma/2}\exp(-\pi \epsilon(\epsilon+U)/2\Gamma)$. The
competition between extended Kondo and local-singlet states occurs
rather generally in systems of coupled quantum impurities
\cite{aguado00,zitko06,zitko06b}. 

A qubit pair represented by two electrons in a DQD and in the contact
with the leads acting as a fermionic bath can not be described by a
pure state. In the case of mixed states of qubit pairs as studied here
the concurrence is related to the reduced density matrix of the DQD
subsystem as in the previous case of delocalized electrons pairs. If
$t_{1(2)}/U$ is not small the electrons can fluctuate between the dots
(and to the leads) and such charge fluctuations introduce additional
states with zero or double occupancy of individual dots
\cite{schliemann01,zanardi02}.  As pointed out by Zanardi in the case
of simple Hubbard dimer \cite{zanardi02} the entanglement is not
related only to spin but also to charge degrees of freedom which
emerge when repulsion between electrons is weak or moderate.

For systems with strong electron-electron repulsion considered here,
charge fluctuations are suppressed and the states with single
occupancy -- the spin-qubits -- dominate and the concept of
spin-entanglement quantified with concurrence can be applied.  We
again use spin-projected density matrix and consider only entanglement
corresponding to spin degrees of freedom. Due to doubly (or zero)
occupied states arising from charge fluctuation on the dots (caused by
tunneling between the dots A and B or due to the exchange with the
electrons in the leads), the reduced density matrix has to be
renormalized. The probability that at the measurement of entanglement
there is precisely one electron on each of the dots is less than
unity, $P_{11}<1$, and the spin-concurrence is then given with

\begin{equation}
C_{11}=C/P_{11},\label{cnorm}
\end{equation}
where $P_{11}=P_{\uparrow\downarrow}+P_{\parallel}$, and
$P_{\uparrow\downarrow} =\langle P^\uparrow_\mathrm{A}
P^\downarrow_\mathrm{B} + P^\downarrow_\mathrm{A}
P^\uparrow_\mathrm{B}\rangle $, $P_{\parallel} = \langle
P^\uparrow_\mathrm{A} P^\uparrow_\mathrm{B} + P^\downarrow_\mathrm{A}
P^\downarrow_\mathrm{B}\rangle$ are probabilities for antiparallel and
parallel spin alignment, respectively. Such a procedure corresponds to
the measurement apparatus which would only discern spins and ignore
all cases whenever no electron, or a electron pair would appear at one
of the detectors at sites A or B.

\begin{figure}
\center{\resizebox{0.78\columnwidth}{!}{%
\includegraphics{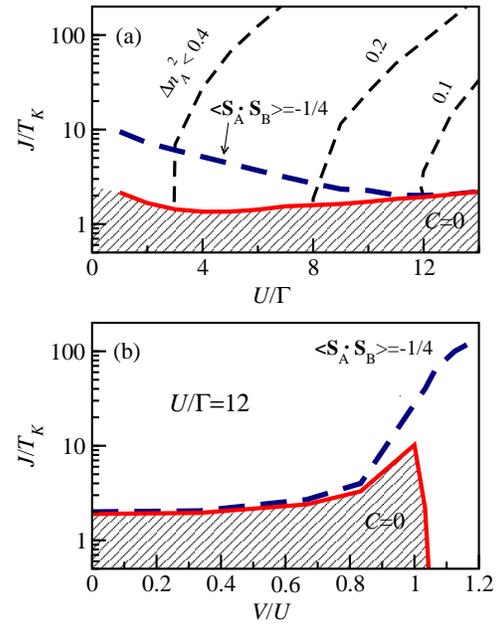} }}
\caption{ (a) Charge fluctuations of the domain A, $\Delta
  n_A^2=\langle n_A^2\rangle-\langle n_A\rangle^2$ (short-dashed),
  $\langle \textrm{\bf S}_A\cdot \textrm{\bf S}_B\rangle=-1/4$
  (long-dashed) and $C=0$ (full) in the $(U/\Gamma,J/T_K)$ plane. (b)
  $\langle \textrm{\bf S}_A\cdot \textrm{\bf S}_B\rangle=-1/4$
  (long-dashed) and $C=0$ (full) in the $(V/U,J/T_K)$ plane. }
\label{fig7}       
\end{figure}

Here we present results for the zero temperature limit obtained by the
projection operator method as in
Refs. \cite{rejec03,mravlje05,mravlje06,mravlje07,epjb08}. The temperature
dependence of the concurrence for the case of $V=0$ is given in Ref.~
\cite{rmzb06}. Expectation values $\langle ...\rangle$ in the
concurrence formula Eq.~(\ref{eq:cmax}) are now calculated using the
ground state therefore $\langle
S_{\mathrm{A}}^{+}S_{\mathrm{B}}^{+}\rangle=0$ and $C_{\parallel}<0$.
We consider the particle-hole symmetric point with $n=2$ and
$\epsilon+U/2+V=0$.

Qualitatively, the concurrence is significant whenever enhanced
spin-spin correlations indicate inter-dot singlet formation. As shown
in Fig.~\ref{fig6}(a) for $U/\Gamma=12$ and $\Gamma/t_0=0.1$, the
correlation function $\langle {\bf S}_\mathrm{A} \cdot {\bf
S}_\mathrm{B} \rangle$ tends to $-3/4$ for $J$ large enough to
suppress the formation of Kondo singlets, but still $J/U\ll1$, that
local charge fluctuations are sufficiently suppressed.  In particular,
the local dot-dot singlet is formed whenever singlet-triplet splitting
superexchange energy $J>J_{c}\sim 2T_{K}$.  With increasing $V\to U$,
and above $U$, the probability for singly occupied spin states,
$P_{11}=P_{\uparrow\downarrow}+P_{\parallel}$ is significantly
reduced, Fig.~\ref{fig6}(b), which also leads to reduced spin-spin
correlation.

The entanglement between qubits quantified by concurrence is small in
the regime where the Kondo effect determines the ground state. The
Kondo screening transfers the entanglement between localized electrons
to the mutual entanglement of localized and the conducting electrons
\cite{adam06}. For $V\sim U$ the Kondo temperature is enhanced and the
corresponding Kondo ground state is competitive towards the localized
singlet state. In Fig.~\ref{fig7}(a) we present a phase diagram of
regions with finite and zero entanglement, separated approximately by
the $J\sim 2T_{K}$ line. In the strong coupling limit the boundary
coincides with the $\langle \textrm{\bf S}_A\cdot \textrm{\bf
S}_B\rangle=-1/4$ dashed line. For $V>U+T_K$ the Kondo screening is
inhibited and the concurrence is increased,
Fig.~\ref{fig7}(b). However, the probability for doubly occupied sites
is not small in this regime and the concept of spin-qubits is not
appropriate there \cite{mravlje05}.

\section{Conclusion} \label{sec5}
In this work we showed that the applicability of the concurrence, an
entanglement measure originally restricted to two distinguishable
particles (spins), can be extended to two regions -- measurement
domains -- occupied by two indistinguishable particles
(electrons). The proposed approach allows for the analysis of
entanglement in a variety of realistic problems, from scattering of
flying and static qubits, the former being represented as wave packets
with a finite energy resolution, to the time evolution of static
qubits due to electron-electron interaction or externally applied
fields. A generalization to systems described as a mixed state and
containing more than two electrons is possible. As an example, a
double quantum dot system occupied, on average, by two interacting
electrons is presented.

We acknowledge support from the Slovenian Research Agency under
contract Pl-0044.

\providecommand{\WileyBibTextsc}{}
\let\textsc\WileyBibTextsc
\providecommand{\othercit}{}
\providecommand{\jr}[1]{#1}
\providecommand{\etal}{~et~al.}

%








\begin{thebibliography}{[10]}

\bibitem{einstein35} A. Einstein, B. Podolsky, and N. Rosen, Phys. Rev. {\bf 47}, 777 (1935).
%
\bibitem{nielsen01}M. A. Nielsen and I. A. Chuang, \emph{Quantum
Information and Quantum Computation} (Cambridge University Press,
Cambridge, 2001).
%
\bibitem{osterloh02}A. Osterloh, L. Amico, G. Falci, and R. Fazio,
Nature \textbf{416}, 608 (2002); S. El Shawish, A. Ram{\v s}ak, and J. Bon{\v c}a, Phys. Rev. B {\bf 75}, 205442 (2007).
%
\bibitem{criteria} D.~P.~DiVincenzo,  {\it Mesoscopic Electron Transport,
NATO Advanced Studies Institute, Series E: Applied Science}, edited by L.
Kouwenhoven, G. Sch{\" o}n, and L. Sohn (Kluwer Academic, Dordrecht, 1997);
cond-mat/9612126.
%
\bibitem{bennett96}C.~H. Bennett, H.~J. Bernstein, S. Popescu, and B. Schumacher,
Phys. Rev. A \textbf{53}, 2046 (1996); C.~H. Bennett, D.~P. DiVincenzo,
J.~A. Smolin, and W.K. Wootters, \emph{ibid.} \textbf{54}, 3824 (1996).
%
\bibitem{hill97} S. Hill and W.~K. Wootters,
Phys. Rev. Lett. \textbf{78}, 5022 (1997).
%
\bibitem{vedral97} V. Vedral, M.~B. Plenio, M.~A. Rippin, and P.~L. Knight,
Phys. Rev. Lett. \textbf{78}, 2275 (1997).
%
\bibitem{wootters98} W. K. Wootters, Phys. Rev. Lett. \textbf{80}, 2245 (1998).
%
\bibitem{schliemann01}J. Schliemann, D. Loss, and A.~H. MacDonald, Phys. Rev. B
\textbf{63}, 085311 (2001); J. Schliemann, J.~I. Cirac, M. Ku{\' s},
M. Lewenstein, and D. Loss,  Phys. Rev. A \textbf{64}, 022303 (2001).
%
\bibitem{ghirardi04}G.C. Ghirardi and L. Marinatto, Phys. Rev. A
\textbf{70}, 012109 (2004);
K. Eckert, J. Schliemann, G. Brus, and M. Lewenstein, Ann.
Phys. \textbf{299}, 88 (2002);
J.~R. Gittings and A. J. Fisher, Phys. Rev. A
\textbf{66} 032305 (2002).
%
\bibitem{zanardi02}P. Zanardi, Phys. Rev. A \textbf{65}, 042101 (2002).
%
\bibitem{vedral03}V. Vedral, Cent. Eur. J. Phys. \textbf{2}, 289 (2003);
D. Cavalcanti, M.~F. Santos, M.~O. TerraCunha, C. Lunkes, V. Vedral,
Phys. Rev. A \textbf{72}, 062307 (2005).
%
\bibitem{divincenzo05} D.~P.~DiVincenzo, Science \textbf{309}, 2173 (2005).
%
\bibitem{coish06} W.~A.~Coish~and~D.~Loss, arXiv:cond-mat/0606550.
%
\bibitem{elzerman03} J.~M.~Elzerman, R.~Hanson, J.~S.~Greidanus,
L.~H.~Willems van Beveren, S.~DeFranceschi, L.~M.~K.~Vandersypen, S.~Tarucha,
and L.~P.~Kouwenhoven, Phys. Rev. B~\textbf{67}, 161308 (2003).
%
\bibitem{chen04} J.~C. Chen, A.~M. Chang, and M.~R. Melloch,
Phys. Rev. Lett. \textbf{92}, 176801 (2004).
\bibitem{hatano05} T. Hatano, M. Stopa, and S. Tarucha, Science
\textbf{309}, 268 (2005).
\bibitem{petta05}
J.~R.~Petta, A.~C.~Johnson, J.~M.~Taylor, E.~A.~Laird,
A.~Yacoby, M.~D.~Lukin, C.~M.~Marcus, M.~P.~Hanson, and~A.~C.~Gossard,
Science \textbf{309}, 2180 (2005).
%
%
\bibitem{ramsak06} A. Ram{\v s}ak, I. Sega, and J.H. Jefferson, Phys. Rev. A {\bf 74}, 010304(R) (2006).
%
\bibitem{notedifference} Note that indexing of states here differs from Ref.~ \cite{ramsak06} where $i=1, ...N$ and $j=1, ...N$, but $s\equiv\uparrow$ and $s'\equiv\downarrow$.
%
\bibitem{jrr05}J.H. Jefferson, A. Ram\v{s}ak, and T. Rejec, Europhys. Lett. {\bf 75}, 764 (2006).
%
\bibitem{gunlycke05}D. Gunlycke, J.H. Jefferson, T. Rejec, A. Ram\v{s}ak, D.G. Pettifor,
and G.A.D. Briggs, J. Phys.: Condens. Matter {\bf 18}, S851 (2006). 
%
\bibitem{giavaras06}G. Giavaras, J.H. Jefferson, A. Ram\v{s}ak, T.P. Spiller, and C.
Lambert, Phys. Rev. B {\bf 74},  195341 (2006); 
M. Habgood, J.H. Jefferson, A. Ram{\v s}ak, D.G. Pettifor, and G.A.D. Briggs, 
Phys. Rev. B {\bf 77}, 075337 (2008).
%
\bibitem{wires} T. Rejec, A. Ram{\v s}ak, and
J.H. Jefferson, J. phys., Condens. matter {\bf 12}, L233 (2000); Phys. Rev. B 65, 235301 (2002).
%
\bibitem{lieb}E.H. Lieb and F.Y. Wu, Phys. Rev. Lett. \textbf{25}, 543 (1968). 
%
\bibitem{jones89}B.A.~Jones and C.M.~Varma, Phys.~Rev.~B \textbf{40}, 324 (1989).
%
\bibitem{aguado00}R.~Aguado and D.~C.~Langreth, Phys.~Rev.~Lett. \textbf{85}, 1946 (2000);
T.~Aono and M.~Eto, Phys.~Rev.~B \textbf{63}, 125327 (2001); R.~Lopez, R.~Aguado, and G. Platero, 
Phys.~Rev.~Lett. \textbf{89}, 136802 (2002); W.~Izumida, O.~Sakai, and Y.~Shimizu, 
Physica B \textbf{259-261}, 215 (1999).

\bibitem{zitko06}R.~\v{Z}itko, J.~Bon\v{c}a, A.~Ram\v{s}ak, and T.~Rejec,
Phys.~Rev.~B \textbf{73}, 153307 (2006).

\bibitem{zitko06b}  R. \v{Z}itko and J. Bon\v{c}a, Phys. Rev. 
B \textbf{73}, 035332 (2006);  {\it ibid.} {\bf 74}, 045312 (2006).


\bibitem{rejec03} T. Rejec and A. Ram\v{s}ak, Phys. Rev. B {\bf 68},
  035342 (2003); {\it ibid.} {\bf 68}, 033306 (2003).
\bibitem{mravlje05} J. Mravlje, A. Ram\v{s}ak and T. Rejec,
  Phys. Rev. B {\bf 72}, 121403(R) (2005); {\it ibid.} {\bf 73}, 241305 (2006).
\bibitem{mravlje06}
\textsc{J.~Mravlje},  \textsc{A.~Ram\v{s}ak},  and  \textsc{T.~Rejec},
\jr{Phys. Rev. B} \textbf{74}, 205320 (2006).
\bibitem{mravlje07}
\textsc{J.~Mravlje},  \textsc{A.~Ram\v{s}ak},  and
 \textsc{R.~\v{Z}itko},
\jr{Physica B} \textbf{403}, 1484 (2008).

\bibitem{epjb08} A. Ram{\v s}ak and J. Mravlje,  Eur. Phys. J. B  {\bf 61}, 419 (2008).
  
\bibitem{rmzb06}A. Ram{\v s}ak, J. Mravlje, R. {\v Z}itko, and
J. Bon{\v c}a, Phys. Rev. B {\bf 74}, 241305(R) (2006).
%
%

\bibitem{adam06} A. Rycerz, Eur. Phys. J. B {\bf 52}, 291 (2006).
%
%
\end{thebibliography}
\end{document}